\documentclass[twocolumn, 10pt]{IEEEtran}
\IEEEoverridecommandlockouts
\usepackage{multirow}
\usepackage{tabularx}
\usepackage{url}
\usepackage{array}
\usepackage{hyperref}
\newcolumntype{P}[1]{>{\centering\arraybackslash}p{#1}}
\newcolumntype{M}[1]{>{\centering\arraybackslash}m{#1}}
\newcolumntype{B}[1]{>{\centering\arraybackslash}b{#1}}
\usepackage{amsmath,amssymb,amsfonts}
\usepackage{algorithm}
\usepackage{algorithmic}
\usepackage{graphicx}
\usepackage{textcomp}
\usepackage{gensymb}
\usepackage{hhline}
\usepackage{tikz}
\usepackage{xcolor}
\usepackage{cite}
\usepackage[utf8]{inputenc}
\usepackage[english]{babel}
\usepackage{colortbl}
\definecolor{LightGray}{gray}{0.85}
\definecolor{DeepGray}{gray}{0.25}
\usepackage{amsthm}

\theoremstyle{definition}

\theoremstyle{remark}

\usepackage{footmisc}

\usepackage{dsfont}
\usepackage[separate-uncertainty = true,multi-part-units = repeat, load-configurations = abbreviations, binary-units=true]{siunitx}
\usepackage{siunitx}
\usepackage{xcolor}

\usepackage{xcolor}

\usepackage{soul}

\begin{document}

\title{ A  Vision of XR-aided Teleoperation System Towards 5G/B5G }
\author{
\IEEEauthorblockN{Fenghe Hu, Yansha Deng, Hui Zhou,  Tae Hun Jung,  Chan-Byoung Chae, and A. Hamid Aghvami}

\thanks{F. Hu, Y. Deng, H. Zhou, and A. H. Aghvami, are with  King's College London, UK (E-mail: \href{mailto:fenghe.hu, yansha.deng, hui.zhou, hamid.aghvami@kcl.ac.uk}{fenghe.hu, yansha.deng, hui.zhou, hamid.aghvami@kcl.ac.uk})(Co-Corresponding author: Yansha Deng).}
\thanks{T. H. Jung and C.-B. Chae are with  Yonsei University, Korea (Email: \href{mailto:taehun.jung, cbchae@yonsei.ac.kr}{taehun.jung, cbchae@yonsei.ac.kr})(Co-Corresponding author: Chan-Byoung Chae).}
}

\maketitle



\begin{abstract}
    Extended Reality (XR)-aided teleoperation has shown its potential in improving operating efficiency in mission-critical, rich-information and complex scenarios. The multi-sensory XR devices introduce several new types of traffic with unique quality-of-service (QoS) requirements, which are usually defined by three measures—human perception, corresponding sensors, and present devices. To fulfil these requirements,  cellular-supported wireless connectivity can be a promising solution that can largely benefit the Robot-to-XR and the  XR-to-Robot links. In this article, we present \textcolor{black}{industrial and piloting use cases and identify} the service bottleneck of each case. We then cover the QoS of \textcolor{black}{Robot-XR and XR-Robot links} by summarizing the sensors’ parameters and processing procedures. To realise these use cases, we introduce potential solutions for each case with cellular connections. Finally, we build testbeds to investigate the effectiveness of supporting our proposed links using current wireless topologies.
\end{abstract}

\section{Introduction}
Intelligent robot  systems  are  everywhere  from  daily  life  to industry manufacturing. Although engineers are building and improving robots with high degree of autonomy to handle complex tasks. Still, when it comes to unstructured tasks or tasks involving uncertainty, an indispensable element is human knowledge. However, the lack of spatial perception largely limit the operating efficiency in conventional video-based teleoperation system. Motivated by this, researchers have raised their attention to extended reality (XR)-aided teleoperation, which is designed to integrate well-registered and highly realistic 3D virtual objects/environments into an teleoperation environment and improve operating efficiency~\cite{Makhataeva2020,schmalstieg2016augmented}. By visualizing and interacting with virtual objects, XR enriches the information for manipulators. XR can thus facilitate remote robotic control by providing a ``transparent" human-machine interface through which human problem-solving and manipulative skills are conveyed to remote environments. The benefits of XR have been exploited in piloting and industrial robots applications~\cite{Makhataeva2020,HIETANEN2020101891}.

\begin{figure*}
    \centering
    \includegraphics[width=0.95\textwidth]{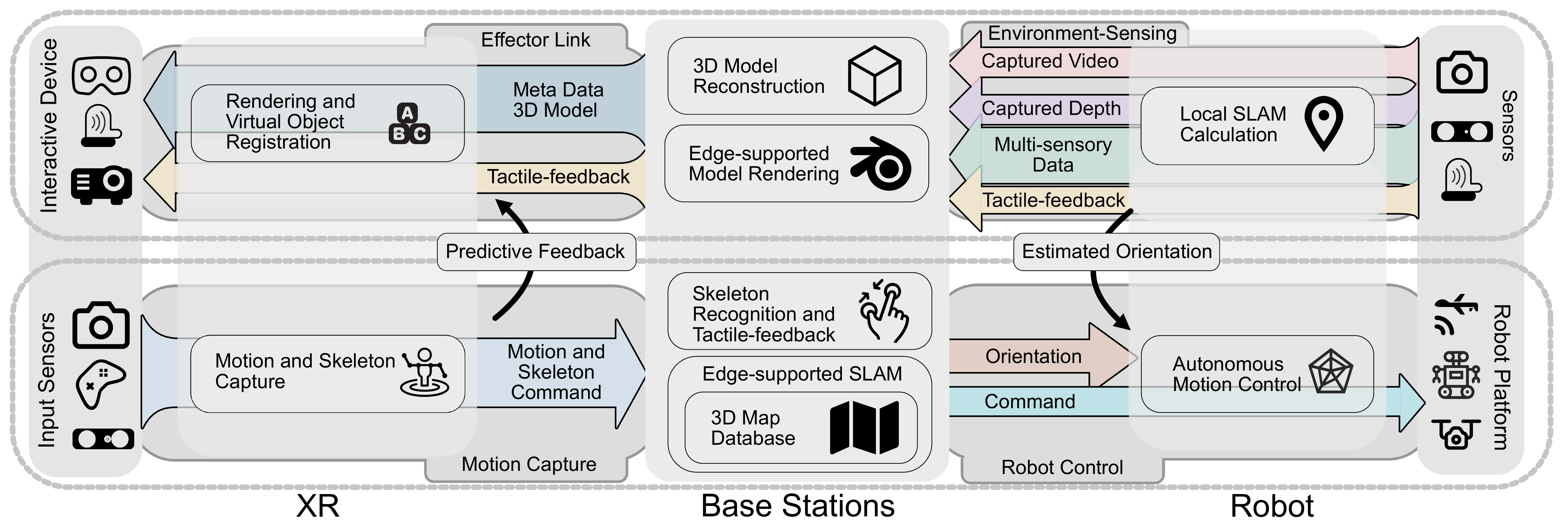}
    \caption{System Architecture of XR-aided Teleoperation with Typical Data Links and Processing Tasks.}
    \label{fig:links}
\end{figure*}

The limitations in wireless communication technology is the primary bottleneck for supporting the demand for real-time transmission and computation of 3D spatial information from XR and control information of robots~\cite{8970173}. With the increasing interests and demands for mobile robots, the existing  wireless  teleoperation  approach fails in support  low-latency and wide geographical area~\cite{Park2018b} as it is  mainly  based on dedicated radio transmission protocols, such as Bluetooth and Wi-Fi. On the other hand, unlike the conventional video-based teleoperation system with direct end-to-end data link, it is necessary to employ a server to handle the reconstruction of 3D environments and the support of  robots’  autonomy.  To support above requirements of XR-aided teleoperation system, a  promising  candidate is 5G/B5G cellular networks~\cite{VRMag}.

Researchers and practitioners today have been focusing on dealing with the challenges to support either XR~ \cite{ExtendXRin5G,VRMag}, aerial-robot control~\cite{NRIIOT}, or ground-robot control~\cite{NRV2X} independently via 5G cellular networks. Current research for XR mainly focus on $360\degree$ video-based transmission with the consideration of users' correlation~\cite{VRMag}. The software-defined network and network function virtualization are also employed to provide differential service for data links in XR or robotics based on link characteristics~\cite{mekikis2019nfv, NRIIOT,NRV2X}, i.e. enhanced Mobile Broadband), Ultra Reliable Low Latency Communications, and massive Machine Type Communications. XR-aided teleoperation system, though, is a closed-loop control system with mixed types of traffic, including depth, 3D XR model data, command \& control data, and multi-sensory data~\cite{8970173}. In this system, the manipulator’s command \& control information is based on the environment observation provided by XR. In many cases, this requires stability and transparency, which can be undermined if transmission or representation errors cause the observed scene from XR to deviate from the desired scene. This calls for stable low latency end-to-end transmission and processing for real-time captured data links~\cite{ExtendXRin5G}. Besides, different stimuli must be aligned in time with tolerable error to enhance transparency. The misalignment and error directly results in deviated observations which can accumulate. In another side, the heavy computation overhead for the processing of structural information calls for computation support. This new mixed environment-correlated traffic and spatial environment information calls for integrated thinking and study of all the consecutive traffic streams and corresponding requirements of an XR-aided teleoperation system as a whole.

The main contribution of this paper are: 1) we first identify \textcolor{black}{industrial and piloting use cases},  with 
 diverse QoS  requirements
 in Section II. We also identify the unique challenges and potential solutions of these two use cases  taking the full advantages of an XR-aided teleoperation system in Section III;
2) we then provide a comprehensive study of the processing tasks and the  data stream characteristics along with its QoS requirements  of the  Robot-to-XR link and the XR-to-Robot link  in an XR-aided teleoperation system supported via 5G/Beyond-5G (B5G) cellular networks in Section IV;
 3) to explore the service quality with current wireless network topologies, we build a virtual reality (VR) teleoperation testbed with force feedback link for industrial via WiFi and Long-Term Evolution (LTE) and a teleoperation unmanned aerial vehicle (UAV) testbed for piloting via LTE and analyse the results in Section V.
  Finally, we conclude the paper in Section VI.

\section{Use Cases}
In this section, we identify \textcolor{black}{industrial and piloting use cases} in XR-aided teleoperation, which have unique traffic requirements.

\subsection{Industrial}
At an industrial site, the workers need a bridge to a rich information environment, where human and robots are collaborating with the support of massive sensors’ data. For example, the workers interact with the robot to handle the moving of heavy products. By overlaying information on the required part of the physical object, XR frees the hands of workers from conventional click-and-move interactions. This kind of interaction largely benefits information interaction efficiency and cooperative robot control by blending virtual and physical spatial information. Such information is usually stored in private edge server due to privacy concern. To support massive robots connected to a MEC server, the high uplink bandwidth and reliability requirement push conventional access technology to its utmost. An intelligent warehouse home to over 500 cooperating robots requires over \SI{10}{\giga bp\second} uplink traffic with video feedback. And their command \& control information requires a service availability of $99.999999\%$ with sub-ms delay and small jitter at \SI{1}{\micro\meter} level to ensure safety and reduce accidental loss~\cite{NRIIOT}.

\subsection{Piloting}
Piloting describes the determination of the curse of ground (wheeled or legged) or aerial vehicles with mobility. Currently, the major operating scheme is the line-of-sight (LoS) control with visual contact or video feedback. In this case, the mismatch between pilots' and robots' coordination cause disorientation, which reduce the efficiency of piloting. An XR-aided teleoperation system can offer an intuitive control experience by immersing pilots in the robots’ coordination system. Moreover, by combining the information in edge/cloud server, i.e., environment structure, map, and co-operating robots, XR can greatly extend a pilot's capability. What is necessary to realise the above benefits is a seamless low-latency connection between the robot and manipulator over a long distance with mobility support. The transmission from robots to XR requires over \SI{50}{\mega bp\second} bandwidth to transmit high-resolution video, depth, and other information gathered by the sensors. The control command requires around \SI{10}{\kilo bp\second}, such as target coordination~\cite{NRV2X}. Moreover, the motion-to-photon delay should be below \SI{50}{\milli\second} to satisfy human perception and ensure safety.

\section{System Architecture and Performance Requirements}
Inspired by use cases, we propose the general architecture of an XR-aided teleoperation system. The system, as illustrated in Section. II and Fig.~\ref{fig:links}, consists of the Robot-to-XR link (shown at the top) and the XR-to-Robot link (shown at the bottom)~\cite{8003431}. In the Robot-to-XR link, robot sensors capture environment information, including video, depth, multi-sensory data, and tactile-feedback. The orientation of robots can also be estimated with information from sensors for fast reacting. Sensors' information is then delivered to the base stations (BS). Supported by the MEC co-locating with the BS, the captured video and depth need to be reconstructed into a 3D model and pre-rendered into metadata. After receiving the rendered model from BS, the XR devices render and register the model with the manipulators’ physical environment. In the XR-to-Robot link, the perceptual devices (e.g., haptic gloves) capture the manipulators’ motion and skeleton, which are then transmitted to BS, where they are recognised as the command to control the robot. With previously captured environment information (e.g., 3D model and texture), it is possible to generate predictive visual or haptic feedback to users before receiving real one. With the edge/cloud-supported simultaneously localization and mapping (SLAM), command and robots’ coordination can be calculated and transmitted to the robot~\cite{schmalstieg2016augmented}. We introduce the physical devices and data rate of each link in the following.

\begin{table*}[t]
\centering
\caption{The specifications of sensors and robot platforms.}
\label{tab:deviceSpec}
\begin{tabular}{|l||c|c|c|c|c|c|}
\hline
\rowcolor{LightGray}
Depth Sensor & Type & bit/point & points/s & FoV & Range & Error \\ \hline
TI AWR1462 & mmWave & / & $0.97$~M & / & \SI{25}{\meter} & \SI{3}{\milli\meter} \\
RealSense D415 & Infrared & $16$ & $27.6$~M &\ang{69}$\times$\ang{42}& \SI{10}{\meter} & / \\ 
Kinect DK & Infrared & $16$ & $7.86$~M &\ang{75}$\times$\ang{65}& \SI{5.4}{\meter} & / \\ 
StereoLabs ZED2 & Binocular & $8$ & $25.3$~M &\ang{110}$\times$\ang{70}& \SI{20}{\meter} & / \\
SLAMTEC A1 & LiDAR & $64$ & $8$~k &\ang{360} 2D& 12m & \SI{2}{\milli\meter} \\ 
Velodyne Hi-Res & LiDAR & $48$ & $60$~k &\ang{360}$\times$\ang{40}& \SI{100}{\meter} & \SI{3}{\centi\meter} \\
Livox Mid-40 & LiDAR & $103$ & $100$~m &\ang{38.4} $\Phi$& \SI{260}{\meter} & \SI{2}{\centi\meter} \\
RealSense L515 & LiDAR & $28$ &  $23.6$~M &\ang{70}$\times$\ang{55}& \SI{6.5}{\meter} & \SI{5}{\milli\meter} \\
\hline
\rowcolor{LightGray}
Robot Platform & Frequency & bit/pose & DoF & \multicolumn{3}{c|}{DoF Details} \\ \hline
DJI Drones & \SI{50}{\hertz} & $232$ & $3$ & \multicolumn{3}{l|}{XYZ Direction}\\
xArm Robot Arm & \SI{100}{\hertz} & $328$ & $6$ & \multicolumn{3}{l|}{6 Motors}\\
GeomagicTOUCH & \SI{1000}{\hertz} & $96$ & $6$ & \multicolumn{3}{l|}{XYZ Direction and Force}\\
Dexmo Glove & \SI{1000}{\hertz} & $1008$ & $21$ & \multicolumn{3}{l|}{16 Instrument 5 Driven}
 \\ \hline
\multicolumn{7}{l}{*Note that there is additional cost for packet head and timestamp.}
\end{tabular}
\end{table*}

\subsection{Robot-to-XR Link}
The Robot-to-XR link mainly carries large size media traffic and tactile feedback, and it contains the environment-sensing link from the robots’ sensors to edge-enabled BS, and the virtual-present link from BS to XR representing devices.

\subsubsection{Environment-sensing Link}
To create the perfect representation of robots’ working environment and support autonomous robot control, the robot is equipped with a wide range of sensors. These include depth sensors, LiDAR sensors, cameras, and multi-sensory information sensors, to capture 3D structure, colour, and other multi-sensory data (force and illumination). The 3D structure is measured by depth cameras, which acquire multi-point distance information across their field-of-view (FoV) from the sensors’ coordination by ranging-image techniques~\cite{schmalstieg2016augmented}. The double-eye depth camera measures the distance through stereophotogrammetry. The stereo setup benefits from its ultra-wide FoV, but it requires a graphics processing unit (GPU) to perform real-time distance calculation, and the resulting accuracy is not guaranteed. Taking StereoLabs ZED2 in Table~ \ref{tab:deviceSpec} as an example, the depth is calculated based on captured two-way video at 100 frame per second with a data rate of \SI{5.9}{\mega\byte/\second} each. The generated depth map has $25.3$~M depth points per second with \SI{32}{\bit} each.

The LiDAR measures the distance by measuring the time-of-flight of a single rotating laser beam point-by-point, which generates depth measurements in sequence. At long distances, the measurement is highly precise with its millimetre depth resolution but it has a low refresh rate. LiDAR usually generates \SIrange[range-units = brackets]{8}{100}{\kilo } points per second, resulting in a data rate around \SI{500}{\kilo\byte/\second} to \SI{1}{\mega\byte/\second} for devices like: SLAMTEC A1, Velodyne Hi-Res, and Livox, as shown in Table I. Another type of LiDAR applies mirrors to perform optical beam-steering for fast scanning, achieving a fast refresh rate with very limited range. Take the L515 LiDAR in Table~\ref{tab:deviceSpec} as an example, it can take $23.6 \text{M}$ points per second in a \SI{6}{\meter} range.

The color and illustration information can be captured by camera, which generates~ \SIrange[range-units = brackets]{5}{70}{\mega bp\second} video data with 1080p to 4K resolution at $60$ frame per second. The multi-sensory information is captured by individual tiny sensors such as IMU, temperature, pressure, and so forth, where the generated information is several bytes per datum. The force is commonly captured by robot arms with force sensors attached to each joint or surface, generating \SI{48}{\mega bp\second} data for each DoF (XYZ data in \SI{1000}{\hertz}).

\subsubsection{Effector Link}
The effector link connects the edge server-enabled BS with the XR, and it delivers the reconstructed and rendered 3D model to the XR devices~\cite{HIETANEN2020101891}.
In MEC/cloud, the data received from the robot is combined and registered into a joint 3D global model for robots' localization and representation at XR’s side. The global model needs to be continually updated, which can also enhance the robots' tracking accuracy and reliability~\cite{schmalstieg2016augmented}. The reconstructed model is rendered to support a perfect representation at the manipulators’ side. Stored in point cloud data, the reconstructed model can be directly rendered into a highly accurate model by drawing points inside the scenes~\cite{8970173}. By controlling the dense of points, the 3D model can be separated into layers with different levels of detail (LoD); this permits scalable and flexible deployment according to different devices or network quality. It is common to have more than one billion points inside a highly detailed model, resulting in a model size of \SI{12}{\giga\byte} and, after compression, \SI{2}{\giga\byte}~\cite{8970173}. The model size can be saved by converting the depth information into a mesh grid, and the colour information into the texture. This method approximates the shape of a real object with meshes, and the quality of the model increases with the number of polygons and resolution of texture. The size of the resulting 3D model varies with environment structure complexity, ranging usually from \SI{10}{\mega\byte} to \SI{500}{\mega\byte}. Then, the model is rendered into metadata or video from the reconstructed model, which can be then combined or up-scaled to XR display with technologies like Nvidia DLSS.

\subsubsection{Human Perception}
To enhance the naturalness and intuitiveness of the system, virtual objects should be presented to match the manipulators’ visual perception and the expectation or to accord with the law of physics with correctly rendered shadows, illumination, occlusion, and generated force feedback.

The see-through or projection displays allow humans to directly see the virtual objects and background environment simultaneously~\cite{schmalstieg2016augmented}. Since both standard XR system and XR-aided teleoperation system aim to immerse users/manipulators into the virtual environment. The virtual objects on display should also satisfy human visual perception and correctly interact with the real environment. To satisfy human visual perception, the spatial resolution of the virtual object should be greater than \SI{60}{pixel/degree} with at least \SI{60}{\hertz} refresh rate~\cite{VRMag}. If the model is rendered into video remotely, it requires 4K resolution to fulfill the $52\degree\times34\degree$ FoV in Hololens2, which consumes \SI{70}{\mega bp\second} bandwidth. The presented model quality is commonly measured by the Peak Signal-to-Noise Ratio, which measures the difference between the generated and presented model. To enable interaction with the environment, it requires information from the operating environment, such as illumination and structure, to be placed precisely and stably without noticeable jitters. To match the physical force feedback to human perception, it requires force input from motors to generate force at each joint. Take the Dexmo gloves as an example; it requires $5$ pair of XYZ data for force direction and strength in \SI{1000}{\hertz}, resulting in \SI{80}{\kilo bp\second} traffic. Here, the force feedback should satisfy Weber’s law of just noticeable differences, which gives the maximum tolerable error.

\subsection{XR-to-Robot Link}
For both manipulator and robots, the XR-to-Robot link carries the necessary command \& control information and state information (e.g., location, orientation, and pose), which are critical for maintaining transparency. 

\subsubsection{Motion Capture}
The capture of manipulators' motion is critical for XR interaction, which is different from the conventional move-and-click style. The manipulators (i.e., skilled human) interact with the 3D objects via direct hand/body gestures and voice command, which can be captured from a specially designed joystick, haptic gloves, and cameras. It has been shown that human require at least \SI{13}{\milli\second} to capture a new virtual signal and that for most motion interactions, like pressing a button, \SI{100}{\milli\second} is sufficient. The error for motion transmission is required below $10^{-6}$ for both XR and robot control~\cite{ExtendXRin5G,NRIIOT}.

For sensor-based capture, the manipulator’s motions are tracked by joystick and haptic gloves by attaching tracking sensors or markers onto each degree-of-freedom (DoF) of a skeleton. The data sampling rate for each DoF is widely set as \SI{1000}{\hertz} to reduce the reaction delay, resulting in more than \SI{100}{\kilo bp\second} data based on the number of tracking joints. For the vision-based capture, the sampling rate is around \SIrange[range-units = brackets]{60}{90}{\hertz} due to constraints in processing capabilities and bandwidth~\cite{HIETANEN2020101891}. It is common to combine both schemes to balance between simplicity and accuracy. Taking Apple’s ARKit as an example, skeleton tracking can be done via the mobile device at \SI{60}{\hertz} with more than $20$~DoF. If the tracking is leveraged to MEC/cloud server, it requires at least \SI{7}{\mega bp\second} for each camera with 1080p video feed. 

\subsubsection{Robot Trajectory Tracking}
One fundamental requirement is the simultaneous tracking of robots’ location and trajectory inside the reconstructed 3D model. This requires geometry perception data from the inertial measurement unit (IMU), satellites, camera, and radar. Defined as SLAM, the procedure can be done in a robot with local sensors or satellites. Another method is to match the captured key frames and depth data with historical information. The procedure can be done in a robot with local poses. With limited computation and storage space at robots, it can continuously estimate its trajectory with associated local poses by taking full advantage of MEC/cloud server with their stored reconstructed map and high definition map~\cite{schmalstieg2016augmented}.

\subsubsection{Robot Control}
Robots can be controlled via directly managing each motor or setting a target. The direct control is widely applied in high-precision industrial computer numerical control (CNC) and surgery systems. As shown in Table~\ref{tab:deviceSpec}, the xArm robot arm needs XYZ coordination data input for each motor with \SI{1000}{\hertz} frequency. Such transmission requires low end-to-end (E2E) latency ranging from sub-ms to \SI{10}{\milli\second} with small jitter at \SI{}{\micro\meter} level to meet the critical closed-loop control requirements. It also requires time synchronization and service availability of \SI{99.999999}{\percent} to ensure the accuracy and safety in a large plant, resulting in the departure of one \SI{1200}{\hertz} packet every \SI{}{\milli\second}~\cite{NRIIOT, Park2018b}. Another method is the target setting. Here, to achieve the target the robot control system automatically controls each joint, which can be a robot’s coordination, pose, or grabbing objects. This requires a much lower refresh frequency. Taking the DJI drone as an example, it requires a pair of XYZ coordination inputs in 50Hz to control its movement.

\section{Challenges and Solutions}
To satisfy the aforementioned links' service requirement and make use cases a reality, specific solutions are required to cope with each challenge. Here, we introduce several typical challenges and their corresponding potential solutions.

\begin{figure*}
    \centering
    \includegraphics[width=\textwidth]{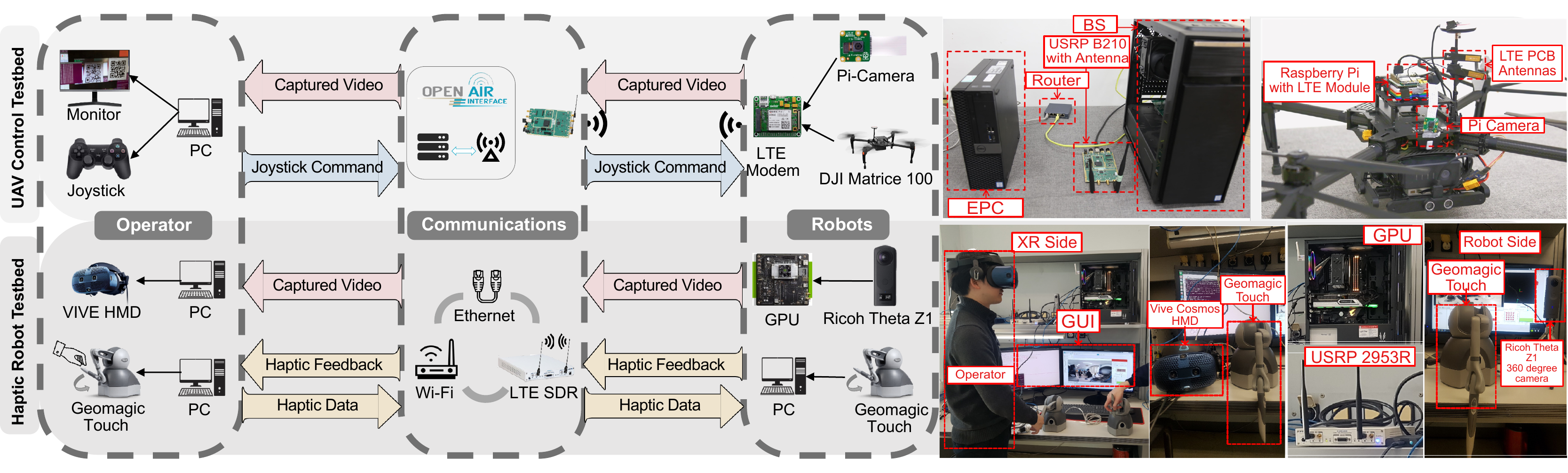}
    \caption{Experiment Environments of UAV Control Testbed and Haptic Robot Testbed.}
    \label{fig:testbed}
\end{figure*}

\subsection{Uplink Capacity}
In the industrial use case, the uplink (Robot-to-XR link) capacity limits the massive deployment of cameras and robots. To cope with the \SI{}{\giga bp\second}-level uplink traffic from hundreds of connected cameras and robots, we propose two potential solutions. The first is dynamic time-division duplex (TDD), which supports high uplink traffic by adjusting the TDD configuration. However, different TDD configurations of adjacent cells lead to cross-link interference (CLI). Therefore, it is possible to employ dynamically control algorithms for the TDD configurations and reduce the CLI. The other potential solution is the utilization of 5G NR operated in an unlicensed spectrum (NR-U). In addition to exploiting its larger bandwidth, the NR-U user can establish dual connectivity with different network topologies, such as WiFi and cellular networks, to further boost uplink capacity. NR-U also adopts synchronized sharing technology, which can utilise Cooperative Muti-Point (CoMP) to improve the uplink capacity. This has been extended to different networks and verified by Qualcomm.

\subsection{3D Coverage}
In the piloting use case, the operation range and reliability of remote control is limited by the coverage of the cellular network at a different altitude. For aerial robots, such as UAV, the communication channel is dominated by LoS links, which causes more severe interference~\cite{EUAV}. Moreover, cellular BSs are tilted downwards and optimized for ground users, while leveraging side-lobe to communicate with aerial robots, resulting in non-contiguous coverage regions. One way to cope with these challenges is to equip the UAVs with directional antennas, which enable the beam steering. By pointing the beam towards their serving base station, the Robot-to-XR link’s performance can be enhanced, while reducing interference. At the BS’s side, interference management and multi-cell coordination can benefit the connection by combining multiple BS into a macro cell and ensuring the connection quality. Apart from that, low-cost, solar-powered small cells with wireless backhaul links (e.g., Nokia F-Cells) can be deployed to enhance the service quality towards the sky. Another way to cope with these challenges is to deploy the satellite network, which is attracting attention as part of 6G to support the communication of drones. For high-altitude aerial robots, they can directly connect to a low-orbit satellite network with low latency support.

\subsection{Mobility with Multi-access Edge Computing}
In both use cases, the heavy computing tasks can be leveraged to multi-access edge servers. Multiple robots can cooperate and jointly sense their environment information with the support of edge servers. The 3D assets can also share between manipulators for rendering. However, it is challenging to keep MEC resources up-to-date with mobile manipulators and robots. During handover between MECs, the synchronization and lack of resources in a target MEC can cause additional delay. This calls for a macro center manager to handle mobility and resource allocation. With a hierarchical structure where edge, fog, and cloud servers coexist, it is possible to offload some non-delay-sensitive tasks to the upper layer and reduce the potential handover between servers. However, the design of an offloading algorithm must account for the characteristics of the tasks~\cite{EDGE}. With dependent tasks, which require historical data or results from other tasks, the overall delay is determined by the slowest task. It also consumes additional time to switch results and historical data between MECs. Furthermore, with different parallelism of the tasks, the required computation resources for tasks is not infinitely divisible and the additional cost for parallel processing should also be taken into account. For example, it is possible to render individual objects in parallel. But it requires additional intermediate processing nodes to composite the rendering result from other nodes.

\subsection{Fast Changing Environment}
In both use cases, it is common to leverage the network optimization decision to intelligent algorithms \cite{mekikis2019nfv}. The network can offer guaranteed differentiated services for data links with varies characteristics in the same network infrastructure. However, due to the network and computation delay, the decision can expire even as it is made due to the fast-changing environment (e.g., delayed position or channel quality). The inconsistency of the network’s current state and captured state can reduce the effectiveness of algorithms. 
To cope with this challenge, it requires both a low-delay connection for network parameters transmission and specially designed control algorithms, which take into consideration the potential environment change. One possible solution is concurrently observing and executing the decision in parallel while adjusting the action as soon as possible if the environment changes. It can significantly improve the performance of robot control in grasping tasks~\cite{ThinkGoogle}.

\begin{table*}[t]
\centering
\caption{Communication Latency for Different Links in UAV Teleoperation.}
\label{tab:command_and_video}
\begin{tabular}{|c|c|c|c|c|c|c|c|c|c|c|c|c|c|c|c|}
\hline
\multicolumn{1}{|c|}{\cellcolor{LightGray}} & \multicolumn{15}{c|}{\cellcolor{LightGray}Control \& Command} \\ \hhline{>{\arrayrulecolor{LightGray}}-|>{\arrayrulecolor{black}}---------------}
\multicolumn{1}{|c|}{\multirow{-3}{*}{\cellcolor{LightGray}}} & \multicolumn{5}{c|}{\cellcolor{LightGray}\SI{20}{\hertz}} & \multicolumn{5}{c|}{\cellcolor{LightGray}\SI{40}{\hertz}} & \multicolumn{5}{c|}{\cellcolor{LightGray}\SI{60}{\hertz}}  \\ \cline{2-16} 
\multicolumn{1}{|c|}{\multirow{-3}{*}{\cellcolor{LightGray}Latency (\SI{}{\milli\second})}} & min & -std & mean & +std & max & min & -std & mean & +std & max & min & -std & mean & +std & max \\ \hline
\SI{0}{\meter} &15 & 17& 24 &31 & 43&15 & 16& 17 &18 & 63& 1016&1297 & 1327 & 1357&1442 \\ \hline
\SI{1}{\meter} &10 &11 & 18 & 25&37 &14 &14 & 15 &16 &26 &1042 &1300 & 1328 & 1356&1403 \\ \hline
\SI{2}{\meter} &15 &16 & 23 & 30& 53&17 &17 & 18 & 19&25 &1072 & 1301& 1330 & 1359& 1491\\ \hline
\multicolumn{1}{|c|}{\cellcolor{LightGray}} & \multicolumn{15}{c|}{\cellcolor{LightGray}Video} \\ \cline{2-16} 
\hhline{>{\arrayrulecolor{LightGray}}-|>{\arrayrulecolor{black}}---------------}
\multicolumn{1}{|c|}{\multirow{-3}{*}{\cellcolor{LightGray}}} & \multicolumn{5}{c|}{\cellcolor{LightGray}\SI{240}{P}} & \multicolumn{5}{c|}{\cellcolor{LightGray}\SI{480}{P}} & \multicolumn{5}{c|}{\cellcolor{LightGray}\SI{720}{P}}  \\ \cline{2-16} 
\multicolumn{1}{|c|}{\multirow{-3}{*}{\cellcolor{LightGray}Latency (\SI{}{\milli\second})}} & min & -std & mean & +std & max & min & -std & mean & +std & max & min & -std & mean & +std & max \\ \hline
\SI{0}{\meter} &157 & 228& 260 &292 & 355& 663&1131 & 1278 & 1425& 1758&1149 &1463 & 1619 & 1775&2025 \\ \hline
\SI{1}{\meter} &945 &2231 & 3688 &5145 & 7389&574 & 1106& 2461 & 3816&8548 &2519 &3812 & 5411 & 7010& 9260\\ \hline
\SI{2}{\meter}* &575 &1368 & 2996 & 4624&8329 &1010 &1159 & 2660 & 4161& 8437&1292 &1177 & 2849 & 4521&8135 \\ \hline
\multicolumn{16}{l}{*Frame rate is adjusted to lower values automatically to avoid connection breakup.} \\
\end{tabular}
\end{table*}

\begin{table*}[t]
\centering
\caption{Communication Latency with Different Networks in Haptic Robot Cases.}
\label{tab:haptic_and_video}
\begin{tabular}{|c|c|c|c|c|c|c|c|c|c|c|}
\hline
\multicolumn{1}{|c|}{\cellcolor{LightGray}} & \multicolumn{5}{c|}{\cellcolor{LightGray} Haptic} & \multicolumn{5}{c|}{\cellcolor{LightGray} Video} \\ \hhline{>{\arrayrulecolor{LightGray}}-|>{\arrayrulecolor{black}}----------}
\multicolumn{1}{|c|}{\multirow{-2}{*}{\cellcolor{LightGray}Latency (\SI{}{\milli\second})}} & min & -std & mean & +std & max & min & -std & mean & +std & max \\ \hline
Ethernet & 10 & 10 & 13 & 16 & 19 & 140 & 143 & 148 & 153 & 157 \\ \hline
Wi-Fi & 184 & 186 & 200 & 214 & 223 & 656 & 666 & 676 & 686 & 689 \\ \hline
LTE & 204 & 203 & 218 & 233 & 247 & 763 & 764 & 776 & 788 & 794 \\ \hline
\end{tabular}
\end{table*}

\section{Case Study and Testbed}

To exploit the challenges above, we built, with current wireless technologies, a VR-aided haptic testbed with force feedback and a UAV teleoperation testbed for the for the industrial and piloting use case respectively (shown in Fig.~\ref{fig:testbed}). 

In the UAV teleoperation case, the testbed contains an XR-to-Robot link and a Robot-to-XR link between a joystick controller and a DJI Matrice 100 UAV transmitting command \& control signal and low-resolution Raspberry Pi camera video, respectively. The cellular network was built with a software-defined radio platform based on universal software radio peripheral  (USRP), while the network functions for core-network were running on an extra server. For the Robot-to-XR link, the video is fed from the Raspberry Pi camera on UAV to the monitor in operators' side through on-broad LTE modem. For XR-to-Robot link, the manipulator controls the robot via a joystick whose command is sent to UAV through USRP base stations.

Table~\ref{tab:command_and_video} presents the E2E UAV command \& control delay versus various transmission frequency for various flight heights, and E2E real-time video delay with three different video resolutions. We observe that the E2E command \& control delay is almost the same for flight heights \SI{0}{\meter}, \SI{1}{\meter}, and \SI{2}{\meter} with \SI{20}{\hertz} \textcolor{black}{(20 times/second)} and \SI{40}{\hertz}, while the delay increases rapidly to around \SI{1330}{\milli\second} with \SI{60}{\hertz} due to the buffer overflow. We also observe that average E2E video streaming delay for flying UAV ($h=$\SI{1}{\meter}, \SI{2}{\meter}) is always higher than that of stationary UAV ($h=$\SI{0}{\meter}). The reason is the low antenna gain provided by side-lobe of BS antennas.

In the haptic robot case~\cite{jungWCNC2020}, the testbed contains a Robot-to-XR link from the panoramic camera to the controller with one \SI{360}{\degree} VR video stream, and bidirectional XR-Robot links between two Geomagic Touch robot arms with force-feedback. For the Robot-to-XR link, the \SI{360}{\degree} video feed comes from a Ricoh Theta to a Vive Cosmos VR headset, supported by an Nvidia Jetson TX2 edge server for video compression. The video is transmitted in 4K resolution and \SI{30}{\hertz}. For the bidirectional XR-Robot links, the haptic packets were sent in \SI{1000}{\hertz} between two robot arms. In this work, we also evaluate the performance with Ethernet, Wi-Fi, and in-house LTE implemented by USRP connections.

\textcolor{black}{Table~\ref{tab:haptic_and_video} shows the E2E delays of haptic and video communication under Ethernet, Wi-Fi, and LTE network, respectively. We can observe that the E2E delays of both haptic and video communication follow LTE $>$ Wi-Fi $>$ Ethernet. This is because Ethernet offers guaranteed large bandwidth and stable communication, and Wi-Fi (40 MHz) provides larger bandwidth for transmission than LTE (20 MHz). For fair comparisons, in our future work, we will conduct extensive experiments with more general system setup.}

\section{Conclusion}
In this article, we first identified industrial and piloting use cases along with their challenges and potential solutions under the cellular network framework. We then introduce a mixture of data types and processing tasks in the Robot-to-XR link and the XR-to-Robot link of an XR-aided teleoperation system supported via the cellular network, to present the robots’ working environment to the manipulator, and to deliver manipulators' command to robots, respectively. From a link-type perspective,  we presented the quality-of-service requirement of each link fulfilling the hardware specifications and human perceptions. Importantly, we further performed case studies by building two cellular-connected teleoperation platforms. 
Our results demonstrated that current wireless topologies can not fully support the UAV teleoperation and haptic robot cases in terms of latency requirement, which may be alleviated via MEC.
Importantly,  
this work provides a comprehensive view of XR-aided teleoperation systems and opens up new research opportunities for optimized solutions to realize this vision.  


\vskip -2\baselineskip plus -1fil

\begin{IEEEbiographynophoto}{Fenghe Hu}
is currently a Ph.D. student in the Center for Telecommunications  Research (CTR), King's College London. 
\end{IEEEbiographynophoto}

\vskip -2\baselineskip plus -1fil

\begin{IEEEbiographynophoto}{Yansha Deng} 
is currently a  Lecturer  (Assistant  Professor) in the  CTR, King’s College London. Her research interests include machine learning for 5G/B5G wireless networks. 
\end{IEEEbiographynophoto}

\vskip -2\baselineskip plus -1fil

\begin{IEEEbiographynophoto}{Hui Zhou}
is currently a Ph.D. student in the  CTR, King's College London.
\end{IEEEbiographynophoto}

\vskip -2\baselineskip plus -1fil

\begin{IEEEbiographynophoto}{Tae Hun Jung}
is currently a Ph.D. student in the School of Integrated Technology, Yonsei University in Korea. 
\end{IEEEbiographynophoto}

\vskip -2\baselineskip plus -1fil

\begin{IEEEbiographynophoto}{Chan-Byoung Chae}
is an Underwood Distinguished Professor at Yonsei University, Korea.   His research interest includes emerging technologies for 5G/6G and molecular communications. 
\end{IEEEbiographynophoto}

\vskip -2\baselineskip plus -1fil

\begin{IEEEbiographynophoto}{Hamid Aghvami} is
a Professor in Telecommunications Engineering from 1993. He is the founder of the Centre for Telecommunications Research at King’s. 
\end{IEEEbiographynophoto}

\end{document}